
\def \ot {\otimes}

\def \g5{\gamma_5}
\def \gm {\gamma^{\mu}}
\def \ra{\rightarrow}
\def \l4{\Bigl( {\rm Tr}( KK^*)^2-({\rm Tr}KK^*)^2\Bigr)}

\def \k2{{\rm Tr}KK^*}
\def \slash#1{/\kern -6pt#1}
\def \di{\slash{\partial}}

\pageno=0

\def\br{\hfill\break\noindent}

%
\magnification=\magstep1
\vsize=23truecm
\hsize=15.5truecm
\hoffset=.2truecm
\voffset=.2truecm
\parskip=.2truecm

\font\ti=cmbx10 scaled\magstep1\font\eightrm=cmr8
\font\ninerm=cmr9
\def \es{e^{-\kappa \sigma}}
\def \ra{\rightarrow}

\def\br{\hfill\break\noindent}

\def \ot {\otimes}
\def \g5{\gamma_5}

\def \ra{\rightarrow}
\def \l4{\Bigl( {\rm Tr}( KK^*)^2-({\rm Tr}KK^*)^2\Bigr)}

\def \k2{{\rm Tr}KK^*}
\def \slash#1{/\kern -6pt#1}
\def \di{\slash{\partial}}

\def \G11{\Gamma_{11}}
\def\m {\cal M}
\def\n {\cal N}

\def \A {\cal A}

\def \ot {\otimes}
\def \g5{\gamma_5}
\def \ra{\rightarrow}
\def \l4{\Bigl( {\rm Tr}( KK^*)^2-({\rm Tr}KK^*)^2\Bigr)}

\def \k2{{\rm Tr}KK^*}
\def \slash#1{/\kern -6pt#1}
\def \di{\slash{\partial}}

\pageno=1
\nopagenumbers
%
\baselineskip=12truept
\footline={\hfill}
{\hfill ETH/TH/94-33 }
\vskip.2truecm
{\hfill 3 September 1994}

\def\br{\hfill\break\noindent}
\def\ra{\rightarrow}

\def \G11{\Gamma_{11}}
\def\m {\cal M}
\def\n {\cal N}

\def \A {\cal A}

\def \ot {\otimes}
\def \g5{\gamma_5}
\def \ra{\rightarrow}
\def \l4{\Bigl( {\rm Tr}( KK^*)^2-({\rm Tr}KK^*)^2\Bigr)}

\def \k2{{\rm Tr}KK^*}
\def \slash#1{/\kern -6pt#1}
\def \di{\slash{\partial}}


%
%
\footline={\hfill}
\centerline{\ti Review of  Non-Commutative Geometric Methods }
\centerline{\ti Applied to Particle Physics}
\vskip1.2truecm
\centerline{  A. H. Chamseddine \footnote*
{\ninerm Talk presented at Quarks 94, May 1994, Vladimir,
Russia}  }
\vskip.8truecm
\centerline{ Theoretische Physik, ETH, CH 8093 Z\"urich Switzerland}
\vskip1.2truecm

\centerline{\bf Abstract}
\noindent
This is a  brief review where some basic elements of non-commutative geometry
are given. The rules and ingredients that enter in the construction
of the standard model and grand unification models in
non-commutative geometry are summarized.
A connection between some space-time supersymmetric theories
and non-commutative geometry is made. The advantages and
prolems of this direction are discussed.
\vskip1truecm
\noindent
\baselineskip=.65truecm
\footline={\hss\eightrm\folio\hss}
{\bf\noindent 1. Introduction}
\vskip.2truecm
\noindent
There is a continuous search for a consistent unified theory of
all fundamental interactions. The main difficulty lies in
unifying space-time geometry with quantum theory, and this is
intrinsically linked to understanding physics at the planck
scale. Since it is not possible to investigate physics directly
at such high scales, we are mainly guided by considerations of
mathematical consistency, elegance and simplicity.
A central assumption in our forumulation of quantum dynamics
is the manifold structure of space-time where the methods
of commutative differential geometry apply. This assumption is
not made in string theory where the space-time manifold arises
in the limit when loops, in the loop space, shrink to points at
lower energies. Another idea in this direction is the study of
non-commutative spaces as initiated by Connes [1]. At present the
list of non-commutative spaces that has been studied is limited,
and only a handful of such spaces correspond to physical theories
[2-6]. What is encouraging is the fact that the simplest
non-commutative space taken to be the product of a continuous
four-dimensional
Minkowski space times a discrete set of two points, gives the
standard model with the Higgs fields and gauge fields unified [2-5].
There are also other examples where the discrete set consists of
more than two points and where the resultant models are grand
unified models [6]. We shall see that this construction has many
attractive features, but it will also be clear that we are
still far from a final unification picture.

The aim of this review is to give the basic ingredients that
go into the construction of a model based on a non-commutative
space, and to spell out clearly the assumptions made, and
whether improvements are possible.
The emphasis here will be on giving an honest and coherent
picture of where things stand and the important problems
to be solved in order to make further progress.

This review is organized as follows. In section two, we give a
summary of the basic elements of non-commutative geometry.
In section three the steps for constructing the standard model
using non-commutative geometry are listed. In section four,
this is generalized
to grand unified models. In section five we give a connection
between some supersymmetric theories and non-commutative
geometry. In section six the question of gravity in
non-commutative spaces is discussed, and section seven is
the conclusion.
\vskip.15truecm
{\bf \noindent 2. Basic notions of non-commutative geometry }
\vskip.15truecm
\noindent
A smooth manifold,
$M$, can be studied by analyzing the commutative algebra, $C^{\infty}
(M)$, of smooth functions on $M$. In fact, $M$ can be reconstructed
from the structure of $C^{\infty}(M)$. The basic idea in
non-commutative geometry [1] is to define a notion of non-commutative
space in terms of a non-commutative (non-abelian) algebra, ${\A}$,
which  is assumed to be
an \underbar{involutive} algebra. This means that there is an
antilinear operation $^*$ taking $a\in \A$ to $a^* \in \A$, ($a^*$ is
the adjoint of $a$ with $(a\cdot b)^* = b^*\cdot a^*$).
We also assume that $\A$ contains an identity element 1. In
this case one says that $\A$ is a unital, involutive algebra. It
defines a notion of a \underbar{compact}, non-commutative space.

Given a unital, involutive algebra $\A$, one can define an algebra
$$
\Omega^* (\A) \;=\;\displaystyle\mathop{\oplus}_{{\rm
n=0}}^{\infty}\;\rm \Omega^{\rm n} (\A)
$$
as the ``universal, differential algebra'' over $\A$, as follows: One
sets $\Omega^0(\A)=\A$, and defines $\Omega^{\rm n}(\A)$ to be the
linear space given by
$$
\Omega^n ({\A})=\biggl\{ \sum_i a_0^i\; da_1^i\ldots da_n^i \ :
a_j^i\; \epsilon {\A} ,\forall i,j \biggr\} ,\qquad n=1,2,\cdots .
$$
Here $da$ denotes an equivalence class of $a\in\A$, modulo the
following relations:
$$
d(a\cdot b) = (da)\cdot b + a\cdot db,\quad d1=0, \quad d^2=0.
$$
An element of $\Omega^{\rm n}(\A)$ is called a \underbar{form of
degree} $n$. Let $\alpha \in \Omega^{\rm n}(\A)$ and $\beta \in
\Omega^{\rm m}(\A)$. Then one  can define the
product, $\alpha\cdot\beta$, of $\alpha$ with $\beta$, and one
verifies that $\alpha\cdot\beta\in\Omega^{\rm n+m}(\A)$, i.e.
$\alpha\cdot\beta$ is a form of degree $n+m$. With this definition of
a product of forms, $\Omega^*(\A)$ becomes an algebra. Defining
$$
(da)^* = - d (a^*)
$$
one immediately deduces from the definition of $\Omega^{\rm n} (\A)$
and from Leibniz rule that, for $\alpha\in\Omega^{\rm n}(\A)$, $\alpha^*$ is
defined and is again an element of $\Omega^{\rm n}(\A)$.

One-forms play a special role as components of connections on a ``line
bundle'' whose space of sections is given by the algebra $\A$. A
1-form $\rho\in\Omega^1 (\A)$ can be expressed as
$$
\rho = \sum_{i} a^i db^i ,
$$
$a^i, b^i$ in $\A$, and, since $d1=0$, we may impose the condition
that
$$
\sum_{i} a^i b^i = 1,
$$
without loss of generality.

Next, we introduce the notion of a (Dirac) $K$-cycle for
$\A$. Let $h$ be a separable Hilbert space, and let $D$ be a
selfadjoint operator on $h$. We say that $(h,D)$ is a (Dirac)
$K$-cycle for $\A$ iff there exists an involutive representation,
$\pi$, of $\A$ on $h$, i.e., a representation (or antirepresentation)
of $\A$ satisfying $\pi (a^*) = \pi (a)^*$, with the properties that

\item{(i)} $\pi (a)$ and $[D,\pi(a)]$ are bounded operators on $h$,
for all $a\in\A$; and
\item{(ii)} $(D^2+1)^{-1}$ is a compact operator on $h$. A $K$-cycle
$(h,D)$ for $\A$ is said to be $(d,\infty)$-summable iff the trace of
$(D^2+1)^{-p/2}$ exists and is finite, for all $p>d$. A $K$-cycle
$(h,D)$ for $\A$ is said to be \underbar{even} iff there exists a
unitary involution $\Gamma$ on $h$, i.e. a bounded operator on $h$
with $\Gamma^* = \Gamma^{-1} = \Gamma$, such that $[\Gamma,\pi(a)]=0$
for all $a\in\A$, and $\{ \Gamma, D\} = \Gamma D + D\Gamma = 0$.
Otherwise $(h,D)$ is called \underbar{odd}.

Given a $K$-cycle $(h,D)$ for $\A$, we define a representation, $\pi$
of $\Omega^* (\A)$ on $h$ by setting
$$
\eqalign{
\pi \bigl(\sum_{i} &a_0^i da_1^i,\cdots da_n^i\bigr)\cr
&=\;\sum_{i} \pi (a_0^i) \bigl[D, \pi (a_1^i)\bigr] \cdots \bigl[
D,\pi(a_n^i)\bigr],\cr}
$$
for any element $\sum_{i} a_0^i da_1^i \cdots da_n^i \in \Omega^{n}
(\A)$, $n=0,1,2,\cdots$ . We also define the spaces of
\underbar{auxiliary fields}
$$
{\rm Aux}\;=\;{\rm ker }\; \pi + d\; {\rm ker }\; \pi,
$$
where
$$
\eqalign{
{\rm ker} \;\pi = \displaystyle\mathop{\oplus}_{\rm
n=0}^{\infty}\;\bigl\{ \sum_{i} & a_0^i\; da_1^i \cdots da_n^i\;: \cr
&\pi\;\bigl(\sum_{i} a_0^i\; da_1^i \cdots da_n^i\bigr) =
0\bigr\},\cr}
$$
and
$$
\eqalign{
d\;{\rm ker}\;\pi = \displaystyle\mathop{\oplus}_{\rm n=0}^{\infty}\; \bigl\{
\sum_{i}da_0^i\; &da_1^i \cdots da_n^i :\cr
&\pi\;\bigl(\sum_i a_0^i\;da_1^i \cdots da_n^i\bigr) = 0
\bigr\}.\cr}
$$
It follows from Leibniz rule that \ Aux \ is a two-sided ideal in
$\Omega^*(\A)$, and hence $\Omega_D^*(\A) = \Omega^*(\A)$ modulo \ Aux \
is a universal differential algebra. \  If \ $\sum_i a_0^i\;da_1^i\cdots
da_n^i$ $\in \Omega^n (\A)$ \ then
$$
\bigl\{ \sum_i \pi (a_0^i) \bigl[D,\pi (a_1^i)\bigr] \cdots \bigl[
D,\pi (a_n^i)\bigr] + \pi (\alpha) :\quad \alpha\in {\rm Aux}\bigr\}
$$
represents an $n$-form,
$$
\alpha = \sum_i a_0^i\;a_1^i \cdots da_n^i \quad {\rm mod \ Aux},
$$
in $\Omega_D^n (\A)$ as an equivalence class of
bounded operators on the Hilbert space $h$.

We define the integral of a form $\alpha \in \Omega^*(\A )$
over a non-commutative space $\A $ by setting
$$
\int \alpha ={\rm Tr}_w \Bigl( \pi (\alpha )D^{-d} \Bigr)
$$
where ${\rm Tr}_w $ is the Dixmier trace [1]. Alternatively this
integral can be defined by the heat-kernel expression
$$
{\lim}_{\epsilon\ra 0}
{{\rm { tr}}(\pi (\alpha ) e^{-\epsilon \vert D
\vert^2 } ) \over
{\rm { tr}}(e^{-\epsilon \vert D\vert^2 })}
$$
For more details on these
somewhat abstract mathematical notions the reader is referred to
[1,4,7].
\vskip.15truecm
{\bf\noindent 3. The standard model in non-commutative geometry}
\vskip.15truecm
One of the simplest extensions for the structure of space-time
is to take it to be a product of a continuous four-dimensional
manifold times a discrete set of two points [2-5].
The algebra is ${\cal A}={\cal A}_1
\ot {\cal A}_2$ acting on the Hilbert space $h=h_1\ot h_2$, where
${\cal A}_1=C^{\infty}(M)$,  and ${\cal A}_2=
M_2(C)\oplus M_1(C)$ the algebras of $2\times 2$ and $1\times 1$
matrices. The Hilbert space is that of spinors of the form
$L=\pmatrix{l\cr e\cr}$ where $l$ is a doublet and $e$ is a
singlet. The spinor $L$ satisfies the chirality condition
$\g5 \ot \Gamma_1 L=L$, where $\Gamma_1 ={\rm diag}(1_2,-1)$
is the grading operator. This implies that $l=l_L$ is left-handed
and $e=e_R$ is right-handed, and so we can write $l_L=\pmatrix
{\nu_L\cr e_L\cr}.$ The Dirac operator is $D=D_1\ot 1 +\Gamma_1
\ot D_2$, where $D_1=\di $ is the Dirac operator on ${\cal A}_1$
and $D_2 $ is the Dirac operator on ${\cal A}_2$, so that
$$
D_l=\pmatrix{\di \ot 1_2& \g5 M_{12}\ot k\cr
\g5 M_{21}\ot k^* &\di \cr},
$$
where $M_{21}=M_{12}^*$ and $k$ is a family mixing matrix.
The geometry is that of a four-dimensional manifold $M$
times a discrete space of two points. The column $M_{12}$
in $D$, the vev of the Higgs
field,  is
taken to be $M_{12}=\mu\pmatrix{0\cr 1}\equiv H_0$. The elements
$a\in \cal A$ have the representation $a\ra
{\rm diag}(a_1,a_2)$ where $a_1$ and $a_2$ are $2\times 2$
and $1\times 1$ unitary matrix-valued functions, respectively.
The self-adjoint one-form $\rho $ has the representation
$$
\pi_l (\rho )=\pmatrix{A_1\ot 1_3 &\g5 H\ot k\cr
\g5 H^*\ot k^* &A_2\ot 1_3\cr},
$$
where $A_1=\sum_i a_1^i\di b_1^i $, $A_2=\sum_ia_2^i\di b_2^i $
and $H=H_0+\sum_i a_1^iH_0b_2^i$.
To be realistic, the quarks
and the SU(3) gauge group must be introduced.
This can be achieved by taking a bimodule structure relating
two algebras $\cal A$ and $\cal B$ [2], where the algebra $\cal B$
is taken to be $M_1 (C)\oplus M_3(C)$, commuting with the action
of $\cal A$, and the mass matrices in the Dirac operator are taken
to be zero when acting on elements of $\cal B$. Then the one-form
$\eta $ in $\Omega^1 (\cal B)$ has the simple form $\pi_l(\eta )
=B_1{\rm diag}(1_2,1)$, where $B_1$ is a $U(1)$ gauge field associated
with $M_1(C)$.
The quark Hilbert space is that of the
spinor $Q=\pmatrix{u_L\cr d_L\cr d_R \cr u_R}$. The representation
of $a\in \cal A$ is: $a\ra {\rm diag}(a_1,a_2,\overline {a_2})$
where $a_1$ is a $2\times 2$ matrix-valued function and $a_2$
is a complex-valued function. The Dirac operator acting on the
quark Hilbert space is
$$
D_q=\pmatrix{\gamma^{\mu} (\partial_{\mu} +\ldots )\ot 1_2
\ot 1_3 &\g5  \ot M_{12} \ot k{'}&\g5 \ot \tilde{ M_{12}}\ot k{''}\cr
\g5  \ot M_{12}^* \ot k^{'*}&\gamma^{\mu}(\partial_{\mu} +\ldots )
\ot 1_3 &0\cr
\g5 \ot \tilde {M_{12}}^*\ot k^{''*}&0&\gamma^{\mu}
(\partial_{\mu} +\ldots )\ot 1_3\cr}\ot 1_3,
$$
where $k'$ and $k^{''}$ are $3\times 3$ family mixing matrices,
and $\tilde {M_{12}}=\mu \pmatrix{1\cr 0\cr}$.
Then the one-form in $\Omega^1(\cal A)$ has the representation
$$
\pi_q(\rho )=\pmatrix{A_1\ot 1_3& \g5 H\ot k'&\g5 \tilde H\ot
k^{''}\cr \g5 H^*\ot k^{'*}&A_2\ot 1_3&0\cr
\g5 {\tilde H}^*\ot k^{''*}&0&\overline {A_2}\ot 1_3\cr},
$$
where $\tilde H_a=\epsilon_{ab}H^b$.
On the algebra $\cal B$ the Dirac operator has zero mass matrices,
and the one form $\eta $ in $\Omega^1(\cal B) $ has the
representation $\pi_q(\eta )= B_2{\rm diag}(1_2,1,1)$ where
$B_2$ is the gauge field associated with $M_3(C)$.
Imposing the unimodularity condition on the algebras $\cal A$
and $\cal B$ relates the U(1) factors in both algebras [3]:
${\rm tr}(A_1)=0$, $A_2=B_1=-{\rm tr}B_2={i\over 2}g_1B$.
We can then write
$$\eqalign{
A_1&=-{i\over 2}g_2 A^a\sigma_a \cr
B_2&=-{i\over 6}g_1B -{i\over 2}g_3 V^i\lambda_i \cr}
$$
where $g_3$ is the SU(3) guge coupling
constant, and $\sigma^a $ and $\lambda^i $ are the Pauli
and Gell-Mann matrices, respectively. It is tempting to conjecture
that the unimodularity condition is related to the absence of
chiral anomalies.
The fermionic action for the leptons is
$$
<L, (D+\rho +\eta )L>=\int d^4x \sqrt g \Bigl(
\overline L \bigl( D_l +\pi_l (\rho )+\pi_l (\eta )\bigr)L
\Bigr),
$$
and, for the quarks it is
$$
<Q, (D+\rho +\eta )Q>=\int d^4x \sqrt g \Bigl(
\overline Q \bigl( D_q +\pi_q (\rho )+\pi_q (\eta )\bigr)Q
\Bigr),
$$
and these can be easily checked to reproduce the standard
model lepton and quark interactions with the correct hypercharge
assignments.

The bosonic action is the sum of  the squares of the curvature in the
lepton and  quark spaces which are given, respectively, by
$$ \eqalign{
I_l&={\rm Tr}(C_l(\theta_{\rho}+\theta_{\eta})^2 D_l^{-4}) \cr
I_q&={\rm Tr}(C_q(\theta_{\rho}+\theta_{\eta})^2 D_q^{-4})\cr}.
$$
where $\theta =d\rho +\rho^2 $ is the curvature of $\rho  $,
and $C_l$, $C_q$ are constant elements of the algebra.
After projecting out the auxiliary fields, it is possible
to show that the above bosonic action reproduces all the
bosonic interactions of the standard model, and with the
same number of parameters [2,4]. The gauge fields and Higgs fields
are unified in this formalism. Only when $C_l$ and $C_q$
belong to the center of the algebra  one gets fixed
values for the top quark mass and Higgs mass, but these
relations cannot be maintained at the quantum level [8].

The main advantage of following the non-commutative
construction of the standard model is that one gets
a geometrical significance for the Higgs field as well
as a prediction for the nature of the Higgs sector (in
this case one doublet). The distance between the two
copies of the four-manifold is given by the inverse of
the weak scale. At very low energies the two copies
are superimposed on each other, and only at the energy
scale of 100 Gev one can "observe" the splitting of the
two copies through the Higgs interactions.
\vskip.15truecm
{\bf\noindent 4. Unification models }
\vskip.15truecm
\noindent
It is natural to ask whether one can go beyond the standard
model in noncommutative geometry. In particle physics it
is popular to consider larger groups such as $SU(5)$ and $SO(10)$
[9] which contain $SU(3)\times SU(2) \times U(1)$. The main advantage
is that the fermion fields could be unified in one or two group
representations. The nicest example of this is
the $16_s$ spinor representation of
$SO(10)$ which contains all known particles and in addition
a right-handed neutrino that can be used to give a very small
Majorano mass to the left-handed neutrino. The gain in
simplicity does not make the theory more predictive. This
is mainly due to the arbitrariness in the Higgs sector. In
unified models there are many possible Higgs representations
that can do the symmetry breaking of the gauge group down
to $SU(3)\times U(1)$ in a satisfactory way. If the Higgs
sector and the breaking mechanism  could be uniquely determined
then grand unified theories could become more predictive.

As we have seen in the non-commutative construction of the standard
model, the Higgs sector was completely determined, and one hopes
that this feature will continue to be present
in unified theories. We have also associated
the distance between the two copies of the manifold with the
energy scale of the electroweak breaking. For unified theories
the breaking is done at least at two different scales and
for this one must require the discrete group to contain
more than two points [6]. Another starting point is the grouping
of the spinor fields and the Hilbert space associated with
the spinors. For example, if we arrange the leptons in the
form $Q=\pmatrix{q_L \cr q_R \cr}$ where $q=\pmatrix{\nu \cr e\cr}$,
then the corresponding algebra ${\cal A}_2$ will be
$M_2(C)\oplus M_2(C)$. The simplest and most natural possibility
corresponds to a discrete space of four points, and where the
fermions are arranged in the form $Q=\pmatrix{q_L\cr q_R \cr
q_L^c \cr q_R^c \cr}$, and the representation $\pi $ acting
on ${\cal A}$ is given by $\pi (a)={\rm diag}(a_1,a_2, \overline
{a_1},\overline{a_2})$ where $a_1$ and $a_2$ are two by two matrices.
One can convince himself that the resulting model would be
the $SU(2)_L\times SU(2)_R\times U(1)_{B-L} $ with the Higgs
fields in the representations $(2,2),(3,1)+(1,3)$ of $SU(2)_L
\times SU(2)_R$. In case the fermionic representations are
not simple such as the $\overline 5 +10 $ for $SU(5)$, then the
situation becomes complicated because one must specify the action
of elements of the algebra on the spinors in ways which does not
seem to be very natural. In this respect it is preferable to take
fundamental, or spinor representations of the algebra for the
space-time spinors. We can therefore summarize the steps
needed in constructing a non-commutative model.\br
1. Specify
the representation of the fermions.\br
2. Choose the number
of discrete points and the symmetry between them. \br
3. Deduce
the appropriate algebra and the map $\pi $ acting on
the Hilbert space of spinors.\br
4. Write down the Dirac operator
acting on elements of the algebra, and in particular choose
the mass matrices to correspond to a good vacuum of the Higgs
fields.

The above rules will completely fix the fermionic and bosonic
sectors of the model. We note that only certain vacuum expectation
values (vevs) of the Higgs fields are allowed because the potential
is orthogonal to the  auxiliary space, and for a generic vev the
auxiliary space is as big as the algebra itself. To give a concrete
example we take the chiral
space-time spinors $P_+\psi $ to be in the $16_s$ representation
of $SO(10)$, where $P_+$ is the $SO(10)$ chirality operator,
and the number of discrete points to be three [10].
The Hilbert space is taken to be that of the spinor
$\Psi =\pmatrix{P_+\psi \cr P_+\psi \cr P_-\psi^c \cr}$ and
where $\psi^c =BC\overline {\psi}^T $, $C$ being the
charge conjugation matrix and $B$ the $SO(10)$ conjugation
matrix. This ordering will guarantee that the spinors can
acquire masses. The algebra ${\cal A}_2$ is takent to be
$P_+\Bigl( {\rm Cliff}SO(10)\Bigr) P_+$.
Let $\pi_0$ denote the representation of $\A$ on the Hilbert space $h_1
\otimes \tilde{h}_2$ on square-integrable spinors for SO(1,3) $\times$
SO(10), where $\tilde{h}_2 = {\rm C\!\!\!I}\enskip^{32}$ is the 32-dimensional
vector space on which $\A_{\rm 2}$ acts. Let $\bar{\pi}_0$ denote the
anti-representation defined by
$\bar{\pi}_0 (a)\;=\; B\;\overline{\pi_0 (a)}\;B^{-1}$.
We define $\pi (a)$ by setting
$\pi (a)\;=\;\pi_0 (a)\;\oplus\;\pi_0 (a)\;\oplus\;\bar{\pi}_0 (a)$.
We choose the Dirac operator $D$
to be given by
$$
D =\pmatrix{\di \ot 1\ot 1 &\g5 \ot M_{12}\ot K_{12} &
\g5 \ot M_{13} \ot K_{13}\cr
\g5 \ot M_{21}\ot K_{21} &\di \ot 1\ot 1&\g5 \ot M_{23}
\cr\g5 \ot M_{31} \ot K_{31}& \g5\ot M_{32}\ot K_{32}  &
\di \ot 1\ot 1 \cr}
$$
where the $K_{mn}$ are $3\times 3$ family-mixing matrices
commuting with  $\pi ({\A})$. We impose the symmetries
$M_{12}=M_{21}={\m}_0 $, $M_{13}=M_{23}={\n}_0$,
$M_{31}=M_{32}={\n}_0^*$, with ${\m}_0={\m}_0^* $. Similar conditions
are imposed  on the matrices $K_{mn}$.
For $D$ to leave the subspace $h$ invariant, ${\m}_0 $ and
${\n}_0 $ must have the form
$$\eqalign{
{\m}_0 &= P_+\bigl( m_0+i m_0^{IJ}\Gamma_{IJ} +m_0^{IJKL}\Gamma_{IJKL}
\bigr)P_+ \cr
{\n}_0 &=P_+\bigl( n_0^I \Gamma_I +n_0^{IJK}\Gamma_{IJK}
+n_0^{IJKLM}\Gamma_{IJKLM}\bigr) P_- \cr}
$$
where $ \Gamma_{I_1I_2\cdots I_n}={1\over
n!\ } \Gamma_{[I_1}\Gamma_{I_2}\cdots \Gamma_{I_n]}$
are  antisymmetrized products of the gamma
matrices.

Next we define an involutive "representation" $\pi :
\Omega^* ({\A})\Rightarrow B(h) $ of $\Omega^* ({\A})$ by
bounded operators on $h$; ($B(h)$ is the algebra of bounded
operators on $h$).
It is straightforward to compute $\pi (\rho )$ and one gets [10]
$$
\pi (\rho )=\pmatrix{A &\g5 {\m}  K_{12}&
\g5 {\n} K_{13}\cr \g5 {\m} K_{12} &A &
\g5 {\n} K_{23} \cr\g5 {\n}^*  K_{31}&\g5 {\n}^*  K_{32} &
B \overline A B^{-1}\cr}
$$
where the fields $A$, $\m $ and $\n $  are given in terms
of the $a^i$ and $b^i$ by
$$\eqalign{
A&= P_+(\sum_i a^i \di b^i ) P_+ \cr
{\m }+{\m}_0 &= P_+ (\sum_i a^i {\m}_0 b^i)P_+ \cr
{\n }+{\n}_0 &=P_+(\sum_i a^i {\n}_0
B\overline {b^i}B^{-1})P_- \cr}
$$
We can expand these
fields in terms of the $SO(10)$ Clifford algebra.
The self-adjointness condition on
$\pi (\rho ) $ implies, after using the hermiticity of the
$\Gamma_I $ matrices, that all the fields appearing in the expansion
of $A, \m $  are real, because both are self-adjoint,
while those in $\n $ are complex. Equating the action of
$A$ on $\psi $ and $\psi^c $ reduces it to an $SO(10)$
gauge field. The structure of the Higgs fields is completely
determined  to be given by $16_s\times 16_s $ and $16_s\times
\overline{16}_s$. Specifying ${\cal M}_0$ and ${\cal N}_0$
determines the breaking pattern of $SO(10)$. For the
potential to give these vevs as its minimum the auxilary space
should be smaller than the algebra. This imposes severe conditions
on the choices of ${\cal M}_0$ and ${\cal N}_0$.
If the number of discrete points in this model is changed, this
would not change the nature of the Higgs fields, but only the
coefficients of their couplings.
The only modification we can make is to add to
the spinor $\psi $ a singlet spinor so as to give the right-handed
neutrino a Dirac mass. In this case a Higgs field $16_s$ will
also be present. We deduce that the most important advantage
of the non-commutative construction is the prediction of the
Higgs representations once the spinors of the model are specified.
This gives well defined models which can be analyzed in detail.
But there is no reduction in the number of parameters
corresponding to  fixed  Higgs representation and an
admisable breaking pattern. From this
discusion one can deduce that only a very small number of models
can be constructed and for each model the Higgs representation
is fixed, and the symmetry breaking pattern severely restricted.
It will be very interesting to classify these models, and to study
the allowed symmetry breakings, and to investigate the
phenomenology of the promissing cases.
\vskip.15truecm
{\bf\noindent 5. Supersymmetry in non-commutative geometry}
\vskip.15truecm
\noindent
Theories with space-time
supersymmetry has many nice properties which are well
known [11]. It is then tempting to construct non-commutative
actions whose classical part has space-time supersymmetry.
The simplest example is
provided by the $N=1$ super Yang-Mills theory in
four dimensions . The
action is given by [12]:
$$
I =\int d^4x  \bigl(-{1\over 4} F_{\mu \nu}^aF^{\mu \nu a}
+{1\over 2} \overline{\lambda^a }\gm D_{\mu} \lambda^a \bigr),
$$
where $\lambda^a $ is a  Majorana spinor in the adjoint
representation of a gauge group G,
$F_{\mu \nu}^a $ is the field strength of the gauge field
$A_{\mu}^a$ and $D_{\mu }$ is a gauge covariant
derivative. This action is invariant under the
supersymmetry transformations
$$\eqalignno{
\delta \lambda^a &=-{1\over 2} \gamma^{\mu \nu}F_{\mu \nu}^a
\epsilon , \cr
\delta A_{\mu}^a &= \overline{\epsilon }\gamma_{\mu} \lambda^a ,
 \cr }
$$
To reformulate this action  using the methods of
non-commutative geometry [1], we first define the triple
$({\cal A}, h, D )$ where h is the Hilbert space $L^2 (M, \tau,\sqrt
g d^4x)\ot C^n $
of spinors on a four-dimensional spin manifold $M$, ${\cal A}$
is the involutive algebra ${\cal A} =C^{\infty }(M) \ot M_n(C)$
of $n\times n$ matrix valued functions, and $D$ the Dirac operator
$D=\di \ot 1_n$ on $h$. The free part of the fermionic action
is written as ${1\over 2}(\lambda , [\di ,\lambda ])$,
where $(, )$ denotes the scalar product on $L^2(S, \tau ,
\sqrt g d^4x )$ given by
$$
(\psi_1 ,\psi_2) =\int_M \sqrt g d^4x \tau \langle
\psi_1 (x), \psi_2 (x) \rangle ,
$$
where $\tau $ is a normalised trace on ${\cal A}$, and
$\langle , \rangle $ denotes the hermitian structure on
the left module ${\cal E}$ which  will be
taken to be equal to ${\cal A}$. Let $\rho $ be a self-adjoint
element in the space $\Omega^1 ({\cal A})$ of one-forms:
$\rho =\sum_i a^i d b^i $,
Then $\pi (\rho )=\sum a[D,b] $ is equal to $\gm A_{\mu} $
where $A_{\mu} =\sum a\partial_{\mu}b $. Since $\rho $ is
self-adjoint and $\gm $ is antihermitian, then $A_{\mu}^*
=-A_{\mu}$. The curvature of $\rho $ is $\theta =d\rho
+\rho^2 $ where $\theta \in \Omega^2 ({\cal A}) $. A
simple calculation shows that $\pi (d\rho )
=\gamma^{\mu\nu}\partial_{\mu}A_{\nu}+
\sum \partial^{\mu}a \partial_{\mu}b $.
If $\pi (\rho ) \in {\rm Ker (\pi )}$, then
$\pi (d\rho )
=\sum \partial^{\mu }a\partial_{\mu}b =-\sum a \partial^{\mu}\partial_
{\mu} b $,
is an independent scalar function. The choice of $\pi (d\rho )$ in
$\pi (\Omega^2 ({\cal A}))\setminus \pi (d {\rm Ker} \pi
\setminus_{\Omega^1({\cal A})})$ is uniquely determined to be
orthogonal to all auxiliary fields, with respect to the inner
product on $\Omega^2 ({\cal A})$. From this we deduce that,
modulo the auxiliary field (i.e. the kernel of $\pi (d\rho )$ ),
$\pi (\theta ) = \gamma^{\mu\nu} F_{\mu\nu}$. The
Yang-Mills action is
$${1\over 2}{\rm Tr}_w (\theta ^2 D^{-4})=
\int \sqrt g d^4x {\rm Tr} (-{1\over 4} F_{\mu\nu}F^{\mu\nu}),
$$
where ${\rm Tr}_w $ is the Dixmier trace [1].
The interacting fermionic action is
$$
{1\over 2}(\lambda ,[D+\rho ,\lambda]) ={1\over 2}
\int \sqrt g d^4x {\rm Tr}
(\overline{\lambda} \gm [\partial_{\mu} +A_{\mu}, \lambda ]).
$$
The supersymmetry transformation for
$\lambda $ and $\rho $ take the simple form
$$\eqalign{
\delta \lambda &=-\pi (\theta )\epsilon , \cr
\delta \pi (\rho )&=\overline{\epsilon } E_a \lambda (E_a ),\cr}
$$
where $E_a $ is a local orthonormal basis of $\Omega_D^1 ({\cal A})
\equiv \Omega^1 ({\cal A})\setminus ({\rm Ker}\pi +d {\rm Ker}\pi )$.
In our case the basis is $E^a =\gamma^a $.

We next consider the $N=2$ super Yang-Mills action [13]. It is
given by
$$\eqalign{
I&=\int d^4x  \Bigl( -{1\over 4} F_{\mu\nu}^aF^{\mu\nu a}
+{1\over 2} D_{\mu}S^aD^{\mu}S^a +{1\over 2} D_{\mu}P^aD^{\mu}P^a
+\overline{\chi }^a\gm D_{\mu} \chi^a  \cr
&\qquad -if^{abc}\overline{\chi }^a (S^b-i\g5 P^b)\chi^c -{1\over 2}
 \bigl( f^{abc}S^bP^c  \bigr)^2 \Bigr), \cr }
$$
where  $S^a$ and $P^a$ are a scalar and pseudoscalar fields,
and $\chi^a $
is a Dirac spinor, all in the adjoint representation of the
gauge group. This action  is invariant under the
transformations:
$$\eqalign{
\delta A_{\mu}^a &=\overline{\epsilon} \gamma_{\mu}\chi^a
-\overline{\chi^a} \gamma_{\mu}\epsilon ,\cr
\delta P^a&=\overline{\chi^a }\g5 \epsilon -\overline{\epsilon}
\g5 \chi^a ,\cr
\delta S^a&= i(\overline{\chi^a}\epsilon -\overline{\epsilon}\chi^a ),\cr
\delta \chi^a &=\bigl(-{1\over 2} \gamma^{\mu\nu}F_{\mu\nu}^a
-\g5 f^{abc}P^bS^c +i\gm (D_{\mu}S^a-i\g5
D_{\mu}P^a)\bigr)\epsilon . \cr }
$$
{}From our experience with the non-commutative construction
of the standard model, and since  the action contains a
complex scalar field  unified with a gauge field, an
obvious guess is to take the non-commutative space to be
$M_4\times ({\rm two\ points})$, with the algebra
$$
{\cal A}=C^{\infty}(M_4)\otimes M_n(C) \oplus C^{\infty}(M_4)
\otimes M_n(C),
$$
and the Dirac operator
$$
D=\pmatrix{
\di \ot 1_n &i\g5 \ot\phi_0 \cr
-i\g5 \ot \phi_0^* & \di \ot 1_n \cr },
$$
acting on the Hilbert space of spinors of the form
$
\lambda =\pmatrix{ L\chi \cr R\chi \cr}$,
where $L={1\over 2} (1+\g5)$ and $R={1\over 2} (1-\g5)$, and
$\chi $ is a Dirac spinor. Elements of ${\cal A}$ are taken
to be operators of the form $\pmatrix{a&0\cr 0&a}$ where $a $ is
a smooth function on $M_4$ with values in $M_n(C)$. The parameters
$\phi_0 $ are taken to be arbitrary except
for the constraint $ [\phi_0 ,\phi_0^* ] =0 $.
A self-adjoint element $\rho $ in the space $\Omega^1({\cal A}) $
has the representation
$$
\pi (\rho )=\pmatrix{ \gm A_{\mu} &i\g5 \phi \cr -i\g5 \phi^* & \gm
A_{\mu} \cr} ,
$$
where $A_{\mu}=\sum a\partial_{\mu}b $, $\phi +\phi_0 =\sum a
\phi_0 b $ and $\phi^* +\phi_0^* =\sum a \phi_0^* b $.
The fermionic action  can  now be simply written as
$$
{1\over 2}\bigl(\lambda ,[D+\pi (\rho ) ,\lambda ]\bigr)=
{1\over 2}\int \sqrt g d^4x
{\rm Tr} \bigl( \overline{\lambda} [D+\pi (\rho ),\lambda ]
\bigr).
$$
A straightforward calculation [14] shows that the bosonic part of the
noncommutative action is  given by
$$\eqalign{
{1\over 4} {\rm Tr}_w \bigl( \theta^2 D^{-4}\bigr) &=\int \sqrt
g d^4 x {\rm Tr} \bigl( -{1\over 4} F_{\mu\nu}F^{\mu\nu}
-{1\over 2}D_{\mu}\phi D^{\mu}\phi
 +{1\over 8} ([\phi ,\phi^*])^2 \bigr). \cr }
$$
Continuing from Euclidean to Minkowski space and inserting
$\phi =S-iP $, we exactly recover the bosonic part of the
supersymmetric action . The supersymmetry transformations
are now very simple:
$$\eqalign{
\delta \lambda &=-\pi (\theta )\epsilon \cr
\delta \pi (\rho )&= \bigl( \overline{\epsilon }E_i \lambda
-\overline{\lambda }E_i \epsilon \bigr) E_i ,\cr}
$$
where $E_i$ a local orthonormal basis of $\Omega_1 ({\cal A})$,
and $\epsilon $ has the same representation as $\lambda $.
In this case the basis can be taken to be
$E_a= \gamma_a \ot 1_2 $, $E_5= i\g5 \ot \tau_1 $,
$E_6= i\g5 \ot \tau_2 $, where $\tau_1$ and $\tau_2 $ are Pauli matrices.

The same analysis can be repeated for the $N=4$ super
Yang-Mills action [15], and one can show that it admits
a non-commutative construction [14]. When this idea is attempted
with $N=1$ super Yang-Mills coupled to matter, one finds that
this is possible, after some assumptions are made, provided
no general superpotential is taken. The interesting problem
to solve is to find the special kind of superpotentials
compatible with non-commutative geometry.
\vskip.15truecm
{\bf\noindent 6. Gravity in non-commutative geometry }
\vskip.15truecm
\noindent
One of the original motivations for seeking a new
structure of space-time was to include gravity
with the other interactions in a consistent way.
Therefore the question of finding the gravitational
action in non-commutative geometry must be posed.
One can make the construction based on generalizing
the basic notions of Riemannian geometry [16]. For this
one defines the metric as an inner product on
cotangent space. One shows that every $K$ cycle over
${\cal A}$ yields a notion of cotangent bundle associated
with ${\cal A}$ and a Riemannian metric on the cotangent
bundle $\Omega_D^1({\cal A}) $. With the connection
$\nabla $ we define on $\Omega_D^1 (\cal A)$
 the Riemann curvature of $\nabla $  by $R(\nabla )
:=-\nabla^2 $, and the torsion $T(\nabla )=d-m.\nabla $,
where $m$ is the tensor product operator. Requiring
the connection to be unitary and the torsion to vanish
we obtain a Levi-Civita connection. If  $\Omega_D^1 (\cal A)$
is a free, finitely generated module then it admits a basis
$e^A , A=1,2,\ldots N, $
and the connection $\omega^A_{\ B}\in \Omega_D^1 (\cal A)$
is defined
by $\nabla e^A =-\omega^A_{\ B}\ot_{\cal A}e^B $.
Let $T^A\in \Omega_D^2 (\cal A)$ be the components of the
torsion $T(\nabla )$ defined by $T^A =T(\nabla )e^A$. Then
$$
T^A =de^A +\omega^A_{\ B}e^B
$$
Similarly we define $R^A_{\ B}\in \Omega_D^2 (\cal A)$ by
$R(\nabla )e^A =R^A_{\ B}\ot_{\cal A} e^B$. Then
$$
R^A_{\ B}=d\omega^A_{\ B}+\omega^A_{\ C}\omega^C_{\ B}.
$$
The analogue of the Einstein-Hilbert action is
$$\eqalign{
I(\nabla )&:=\kappa^{-2}<R^A_{\ B}e^B,e_A> +\Lambda <1,1>\cr
 &=\kappa^{-2}\int_M {\rm tr} (R^A_{\ B}e^B (e_A)^*)
 +\Lambda \int 1,\cr}
$$
where $\kappa^{-1}$ is the Planck scale.
When this formalism is applied to the product $M_4\times
Z_2 $ one finds that [16]
$$
I(\nabla )=2\int_M (\kappa^{-2}r-2\partial_{\mu}\sigma \partial^{\mu}
\sigma +\Lambda )\sqrt g d^4 x
$$
where $r$ is the scalar curvature of the classical Levi-Civita
connection, and $\sigma $ is a massless
scalar field  couples to the metric of $M$.
To better understand the role of the field $\sigma $ we can
study the coupling of gravity to the Yang-Mills sector [17]. In the
case of the standard model the field $\phi =e^{-\kappa \sigma }$
replaces the electroweak scale. In other words, the vev of the
field $\phi $ determines the electroweak scale.
To determine the $\sigma $ dependence in the Yang-Mills action
of the standard model, we consider the $\sigma $ dependence in
the Dirac operator. For example, the leptonic Dirac operator
is
$$
D_l=\pmatrix{\gamma^a e_a^{\mu} (\partial_{\mu} +\ldots )\ot 1_2
\ot 1_3&\g5 \es \ot M_{12}\ot k\cr
\g5 \es \ot M_{12}^* \ot k^*&\gamma^a e_a^{\mu}(\partial_{\mu} +\ldots
)\ot 1_3\cr}
$$
The bosonic action is invariant  under  rescaling of
the Dirac operator
$D\ra e^{-w}D$, as this implies $g_{\mu\nu}\ra e^{2w}g_{\mu\nu}$ and
$\kappa\sigma \ra \kappa \sigma +w$.
The quantum corrections to the classical potential will depend
on $\sigma $, and the vev of $\sigma $ could be determined
from the minimization equations. This excercise, when applied to
the standard model, gives one extra equation. The minimization
equations could only be solved with a heavy top quark mass
in the region $104\le m_t \le 147 $ Gev , and a Higgs mass
$m_H=1200 $ Gev . But this lies in the region where perturbative
methods fail. Therefore at present one cannot make a prediction for a
signature of  non-commutative geometry when applied to the
standard model.
\vskip.15truecm
{\bf\noindent 7. Conclusions and comments }
\vskip.15truecm
\noindent
We have seen that non-commutative geometry provides a
powerful tool to deal with spaces that could not be dealt
with using the usual methods of differential geometry.
The examples that we have considered so far are based
on the simple geometry of a manifold times a discrete
set of points, and it appears  that to make more
progress we have to
consider space-times which are completely non-commutative.
Since all the data of a non-commutative space are
encrypted in the triplet $({\cal A}, h,D)$, one would
like to determine the characteristics of a physically
good set. Another important point to study is the
search for the  symmetries present in a non-commutative action,
and how to quantize such actions, and whether the new
symmetries, if present, could be maintained at
the quantum level. Another important problem is to
understand better the connection between space-time
and non-commutative geometry, and in particular,
to find out the special proporties of Dirac
operators of supersymmetric theories. The study
of gravitational fields for a non-commutative spaces
has to be studied further, and in particular the
dynamical degrees of freedom present in such theories.
Therefore, although the methods of non-commutative
geometry when applied to simple spaces give rise
to very nice models in particle physics,
many open problems remain. We
hope that the solution to some of these problems
would shed light on the structure of the unified
theory, and make further progress possible.

\vskip.15truecm
{\bf\noindent Acknowledgments}\hfill\break
I would like to thank J\"urg Fr\"ohlich and Giovanni Felder
for an enjoyable collaboration
and Daniel Kastler for stimualting discussions.

\vskip.5truecm
{\bf \noindent References}
\vskip.2truecm
\item{[1]} A. Connes, {\sl Publ. Math. IHES} {\bf 62}, (1983) 44;\br
 A. Connes, in {\sl the interface of mathematics
and particle physics }, Clarendon press, Oxford 1990, Eds
D. Quillen, G. Segal and  S. Tsou.

\item{[2]} A. Connes and J. Lott,{\sl Nucl.Phys.B Proc.Supp.}
{\bf 18B}, (1990) 29, North-Holland, Amsterdam;
{\sl  Proceedings of
1991 Summer Cargese conference} p.53  editors J. Fr\"ohlich
et al (1992),Plenum Pub.;\br
M. Dubois-Violette, R. Kerner, and J. Madore,
{\sl Class. Quant. Grav.} {\bf 6}, (1989) 1709; {\sl J. Math. Phys.}
{\bf 31} (1990) 316;\br
M. Dubois-Violette, {\sl C. R. Acad. Sc. Paris} {\bf 307 I}, (1989)
403;\br
J. Madore, {\sl Mod. Phys. Lett.} {\bf A4},  (1989) 2617.

\item{[3]} D. Kastler, {\sl A detailed account of Alain Connes'
version of the standard model in non-commutative geometry} {\bf I,
II, III}, {\sl Rev. Math. Phys.} {\bf 5}, (1993) 477, and to appear;\br
D. Kastler and T. Sh\"ucker, {\sl Theor. Math. Phys.} {\bf 92},
(1993) 522.

\item{[5]} R. Coquereaux, G. Esposito-Far\'ese and G. Vaillant,
{\sl Nucl. Phys.} {\bf B353}, (1991) 689;\br
R. Coquereaux, G. Esposito-Far\'ese and F. Scheck,
{\sl Int. J. Mod. Phys.} {\bf A7}, (1992) 6555;\br
B. Balakrishna, F. G\"ursey and K. C. Wali, {\sl Phys. Lett.}
{\bf 254B}, (1991) 430;\br
R. Coquereaux, R. Haussling, N. Papadopoulos and F. Scheck,
{\sl Int. J. Mod. Phys. } {\bf A7}, (1992) 2809; \br
A. Sitarz {\sl Phys. Lett.} {\bf B 308}, (1993) 311;\br
H.-G. Ding, H.-Y. Guo, J.-M. Li and Ke Wu, Beijing preprint
ASITP-93-23.

\item{[6]} A.H. Chamseddine, G. Felder and J. Fr\"ohlich,
{\sl Phys. Lett.} {\bf 296B}, (1993) 109; {\sl Nucl. Phys.}
{\bf B395}, (1993) 672.

\item{[7]} A. Connes, {\sl Non-Commutative Geometry}, Academic press,
to be published;\br
D.Kastler and M. Mebkhout, {\sl Lectures on non-commutative
geometry and applications to elementary particle physics},
World Scientific, to be published;\br
A.H. Chamseddine and J. Fr\"ohlich, {\sl Some elements of Connes'
non-commutative geometry and space-time physics}, to appear in
Yang-Festchrift, ed S. Yau.

\item{[8]} E. Alvarez, J.M. Garcia-Bondia and C.P. Martin,
{\sl Phys. Lett} {\bf B306}, (1993) 55.

\item{[9]} H. Georgi and S. Glashow, {\sl Phys. Rev. Lett.}
{\bf 32}, (1974) 438;\br
H. Georgi in {\sl Particles and Fields} {\bf 1974} p.575,
editor C.E. Carlson (AIP, New York, 1975);\br
H. Fritszch and P. Minkowski, {\sl Ann. Phys.} (N.Y) {\bf
93},  (1977) 193.

\item{[10]} A.H. Chamseddine and J. Fr\"ohlich, {\sl Phys. Rev.}
{\bf D50}, (1994) 2893.

\item{[11]} For reviews and references on supersymmetry see:\br
J. Bagger and J. Wess, {\sl Supersymmetry and Supergravity},
Princeton University Press, 1983;\br
P. West, {\sl Introduction to Supersymmetry and Supergravity},
World Scientific, (1986).

\item{[12]} S. Ferrara and B. Zumino, {\sl Nucl. Phys. } {\bf
B79}, (1974) 413;\br
A. Salam and J. Strathdee, {\sl Phys. Rev.} {\bf D11}, (1975) 1521.

\item{[13]} A. Salam and J. Strathdee, {\sl Phys. Lett.} {\bf
51B}, (1974) 353;\br
P. Fayet, {\sl Nucl. Phys.} {\bf B113}, (1976) 135.

\item{[14]} A. H. Chamseddine, {\sl Phys. Lett.} {\bf B332}, (1994)
349.
\item{[15]} L. Brink, J. Schwarz and J. Scherk, {\sl Nucl. Phys. }
{\bf B121}, (1977) 77;\br
F. Gliozzi, J. Scherk and D. Olive, {\sl Nucl. Phys.} {\bf B122},
(1977) 253.

\item{[16]} A.H. Chamseddine, G. Felder and J. Fr\"ohlich,
{\sl Comm.Math.Phys} {\bf 155}, (1993) 205.

\item{[17]} A.H. Chamseddine and J. Fr\"ohlich, {\sl Phys. Lett.}
{\bf B314}, (1993) 308.

\end